# Repository-Aware File Path Retrieval via Fine-Tuned LLMs

Authors: Vasudha Yanuganti, Ishaan Puri, Swapnil Chhatre, Mantinder Singh, Ashok Jallepalli, Hritvik Shrivastava, Pradeep Kumar Sharma
Affiliation: Persistent Systems

**Abstract**—Modern codebases present challenges for developers or AI agents, or AI coding assistants, etc trying to locate relevant source files when answering questions like "How does this feature work?" or "Where is the bug likely introduced?". Traditional code search (e.g., keyword or IR-based) often fails to capture semantic context or cross-file relationships, while large language models (LLMs) excel at understanding natural language but struggle with repository-specific detail[1]. In this paper, we propose a new approach to file path retrieval in code repositories by fine-tuning a strong LLM to directly predict relevant file paths given a natural language query. Our method uses Qwen3-8B model fine-tuned with QLoRA (4-bit Low-Rank Adaptation) and Unsloth optimizations for efficiency. To generate training data, we introduce six novel code-aware strategies that leverage abstract syntax tree (AST) structure and repository content to create realistic question–answer pairs, where each answer is a set of file paths. These strategies span from fine-grained, single-file prompts to hierarchical repo summaries, ensuring diverse coverage of repository knowledge. We fine-tune the LLM on the synthetic QA dataset (covering Python projects Flask, Click, Jinja, FastAPI, and the much larger PyTorch) to teach it to map questions to relevant files. Experiments demonstrate that our fine-tuned model achieves high retrieval accuracy on small-to-medium repositories (exact match up to 91% and recall 93% on held-out queries), significantly outperforming naive single-strategy training. Even on a large-scale codebase (PyTorch, ~4k Python files), the model remains effective (59% recall), highlighting the scalability of the approach. We compare it to traditional retrieval and discuss how multi-level code information enables the LLM to reason about cross-file context. This work presents a first step toward repository-aware code assistants that point developers to relevant source files instead of directly answering, combining the strengths of LLMs and structured code analysis. We conclude with insights on dataset design, the impact of various strategies, limitations (e.g., context length in huge repos), and future directions for integrating retrieval and LLM-based code intelligence.

## Introduction and Motivation

Developers frequently ask questions about a codebase that requires finding *where* certain functionality is implemented, which parts of the code are relevant to a bug or feature, or how different modules interact. Answering such questions often reduces to locating **relevant source files or directories** in a repository. Traditional solutions rely on text-based code search (e.g., grep or BM25-based search indices) or developer knowledge, which can be time-consuming and errorprone. For example, a keyword search might return many irrelevant hits if a term is common, and it fails to capture when the concept is implemented under different names or across multiple files. Classic information retrieval methods like Lucene/BM25 have been used for code search, but they match keywords literally and ignore code semantics. As a result, their precision in code retrieval is limited: for instance, Gu *et al.* found that a Lucene-based code search ranked relevant results much lower compared to a learned code search model[2]. This gap has motivated research into **semantic code search** using machine learning, which represents code and queries in vector spaces to improve relevance.

In recent years, **deep learning and large language models (LLMs)** have transformed code understanding and generation. Models such as *CodeBERT*[3] and *GraphCodeBERT*[4] learn joint embeddings of code and natural language, achieving state-of-theart results on code search benchmarks by capturing semantic and structural information in code. CodeBERT, for example, is a Transformer pre-trained on paired code–documentation data that significantly outperforms earlier methods on natural language code search[3]. GraphCodeBERT extends this by incorporating code's data flow graph, yielding further improvements on tasks like code clone detection and search[4]. These models illustrate the benefit of **structure-aware representations** in retrieving code relevant to a query. Beyond static retrieval models, the emergence of **GPT-style LLMs** (e.g., OpenAI Codex, GPT-4, Code Llama) has enabled more flexible code intelligence. GPT-4 can reason about code and answer high-level questions. However, off-the-shelf LLMs often **lack repository-specific context** and struggle with questions about a specific codebase[1][5]. They may hallucinate filenames or miss relevant modules if those details were not seen in pre-training. As Athale and Vaddina (2025) note, even powerful LLMs can falter on evolving codebases due to outdated or missing context, and naive code search can return contextually irrelevant results[6][7].

Our work addresses this gap by **fine-tuning an LLM to become repository-aware** – that is, to internalize the structure and content of a given codebase so that it can retrieve relevant file paths in response to natural language queries. Instead of generating answers or code, the model's task is constrained to pointing the developer to the **most relevant files** that likely contain the answer. This approach combines the strengths of LLMs (understanding natural language and complex questions) with a targeted retrieval objective. By doing so, we avoid having the model produce free-form explanations (which risk hallucination); it only produces file path references drawn from the repository.

Unlike the conventional use of fine-tuning to expand an LLM's general knowledge or shape its style of generation, we use fine-tuning to bind the model to a specific repository snapshot, so it behaves as a pointer. Given a natural-language query, the model's job is not to explain or generate code broadly, but to select file paths that likely contain the answer. In effect, we convert the model's parameters into a compact, parametric index of the repository. This reframes fine-tuning from open-vocabulary text generation to closed-set, set-valued prediction over the finite set of repository paths. Rather than altering the LLM's broader coding knowledge, our approach grounds it with repositoryspecific factual associations. Compared to RAG or IR, which rely on **external indices at inference time**, our approach **compiles retrieval into the model** yielding single-forward-pass latency and stable, deterministic path predictions.

To train such a model, we need a substantial set of question→file paths examples. Manually labeling which files answer arbitrary dev questions is infeasible. We therefore developed a **pipeline to automatically generate a labeled QA dataset from the codebase itself**. Specifically, we employ Qwen to generate realistic questions about the repository and identify the relevant file(s) as answers. A key insight is that to cover different granularities of questions (from high-level architecture down to specific functions), a single summary or method will not suffice. We design **six complementary strategies** for data generation, each focusing on a different view of the code.

These strategies produce overlapping but complementary sets of QA pairs. Strategy 1 (per-file) and 4 (fine AST) focus on very localized questions; Strategies 2 and 5 produce broader questions that involve multiple files; Strategy 3 is intermediate; Strategy 6 is a fallback for scale. Figure 1 illustrates how these strategies relate, from whole-repo summary to individual file content. By **merging all QA pairs** from S1–S6, we obtain a comprehensive training set that covers a wide range of question types: from "What does class X do?" (answer: single file containing class X) to "How do module A and B interact when handling request Y?" (answer: two or three file paths across those modules).

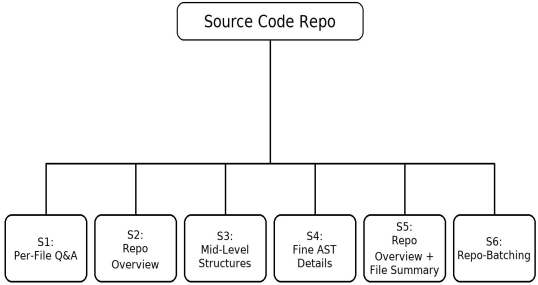

Figure 1: Data Generation Strategies

After constructing the dataset, we fine-tune a large language model to learn this mapping. We use **Qwen38B**, a recently released open-source LLM by Alibaba, chosen for its strong performance and relatively manageable size (8B parameters) for fine-tuning. To efficiently fine-tune on our data, we apply **QLoRA**[21] – an approach that freezes the base model weights and trains low-rank adapter layers on quantized weights
(4-bit), drastically reducing memory requirements[21]. This allowed us to fine-tune on a simple setup of 2 GPUs. We further leveraged *Unsloth* optimizations
(a toolkit for faster training) which provided ~2× speedup by optimizing the training loop (e.g., fused operations and memory utilization)[22]. The model is fine-tuned using the **full repository QA pairs** formatted with a special prompt (described below). Intuitively, the model learns to "pay attention" to certain filenames in the list when certain question patterns appear, effectively internalizing associations between questions and relevant files.

Our contributions are summarized as follows:

- **Novel Problem Formulation:** We tackle the task of *file path retrieval* for natural language queries about a codebase. Rather than generating answers or code, the LLM is trained to output a set of file paths – an approach that, to our knowledge, is new in the context of repository Q&A tools.

- **Automated Dataset Generation with Code-Aware Strategies:** We design a pipeline utilizing six strategies to create a high-quality question→file paths dataset from a repository. This pipeline leverages AST parsing and powerful LLM prompting (Qwen) to cover multiple granularities of questions, enabling the fine-tuned model to reason about both high-level architecture and low-level implementation details.

- **Fine-Tuning a Large Language Model with QLoRA:** We demonstrate that a relatively small LLM (8B) can be fine-tuned to effectively perform retrieval tasks. Using QLoRA and training optimizations, we efficiently train on tens of thousands of synthetic QA pairs using 2 A100 80GB GPUs. The unified prompt format we employ ensures the model's outputs are constrained and factual (file paths from the repo).

- **Experimental Validation on Multiple Repositories:** We evaluate our approach on repositories of varying sizes and domains: *Flask*, *Click*, *Jinja2* (small to medium Python projects, ~50–100 source files each), *FastAPI* (larger, ~1000 Python files), and *PyTorch* (very large, ~19k files, of which ~3.9k are Python). We report metrics (Exact Match and Recall) for the model's retrieval accuracy and analyze how different data generation strategies impact performance. Our fine-tuned models achieve high accuracy on the smaller repos and reasonably good performance on PyTorch, indicating the method's scalability. We also discuss limitations observed, such as the drop in exact match on the largest repo and the challenges of incomplete context in generation.

The remainder of this paper is organized as follows. In Section 2 we review related work in code search and LLM-based code intelligence. Section 3 describes our methodology, including dataset generation and model fine-tuning details. Section 4 presents the experimental setup and quantitative results, with comparisons and discussion. Section 5 outlines the key insights, limitations of our approach, and potential future improvements. Finally, Section 6 concludes the paper.

Table 1: Experiment repo details

| Repository | Total Files | Code Files Used |
|---|---|---|
| Flask | 200 | 94 |
| Click | 167 | 61 |
| Jinja | 133 | 63 |
| FastAPI | 2515 | 1016 |
| PyTorch | 19385 | 3940 |

## Related Work

**Traditional Code Retrieval:** Locating code relevant to a query has long been studied in software engineering. Early and widely used approaches treat code search similar to text document search: indexing code with information-retrieval techniques. Tools like **Lucene/Elasticsearch** (BM25) have been applied to code corpora to allow keyword queries. While fast and simple, pure lexical matching often yields low precision for code search, as code may implement a concept using different terminology than the query. For example, a search for "delete user function" might miss a function remove_account if the names differ. Researchers have proposed augmenting lexical search with symbolic analysis or metadata. Some approaches use **API documentation and code comments** to improve matches or restrict search to function signatures. Nevertheless, lexical methods struggle to understand semantics or relationships across files. As reported by Gu *et al.* (2018), a deep learning model could retrieve relevant code snippets in higher ranks than a conventional search engine[2], underlining the limitations of traditional code search.

**Neural Code Search and Pre-Trained Models:**
The application of deep learning to code search led to significant advances. One line of work focuses on learning embeddings for code and queries in the same vector space so that relevant code is found via nearestneighbor search. **DeepCS (Gu *et al.*, 2018)** was a pioneering neural code search approach that learned joint embeddings of natural language queries and code snippets[23]. It demonstrated improved results over previous techniques like CodeHow and textual search. Following this, large-scale **pre-trained models for code** have emerged. *CodeBERT* (Feng *et al.*, 2020) is a bi-modal Transformer pre-trained on a massive corpus of code and accompanying texts (from CodeSearchNet). It achieved state-of-the-art on natural language code search and code documentation generation[3]. CodeBERT essentially learns the alignment between code and language, enabling semantic search beyond exact token matches. Building on that, *GraphCodeBERT* (Guo *et al.*, 2021) incorporates structural code information (data flow graphs) into the pre-training objective[24][4]. On code retrieval tasks, GraphCodeBERT outperforms CodeBERT by leveraging the code's inherent graph structure, which helps it understand relationships like variable flow and thereby retrieve code in a more semantics-aware way[25]. Other notable models include **UniXcoder** (Guo *et al.*, 2022), which unified code and text modalities and utilized both AST and documentation information in pre-training[26]. These models can be fine-tuned for code search, QA, or summarization tasks and have largely become the foundation for modern code intelligence research.

**Graph-Based and Retrieval-Augmented Methods:** As codebases grow and complexity, capturing relationships across files becomes important. Recent works have introduced *graph-based retrieval* and *knowledge graphs* for code. For example, Athale and Vaddina (2025) propose representing a repository as a **knowledge graph** of code entities and their relations to improve retrieval for assisting code generation[27]. By querying this graph, their approach finds contextually relevant code pieces that a vanilla search might miss, thereby providing the language model with more accurate context. In the domain of LLM-assisted coding, retrieval augmentation has gained traction. **GraphCoder** (Liu *et al.*, 2024) is a framework for repository-level code *completion* that augments an LLM with repository-specific code context via a graphbased retrieval process[28]. GraphCoder constructs a *Code Context Graph* (CCG) of the repository (capturing control-flow and dependence relations around the code to complete) and performs a coarse-to-fine retrieval of similar contexts from the repo[29]. These retrieved code snippets are then provided to the LLM to guide completion. This approach improved exact match accuracy of code completion while being efficient in time and memory[30]. Although GraphCoder targets code generation rather than Q&A, it exemplifies the benefit of combining LLMs with structured retrieval: the LLM's general knowledge is grounded with relevant pieces of the specific repo, overcoming the lack of repository-specific knowledge[31][5]. Our work is similar in spirit – we also recognize the value of repository-specific context – but differs in execution. Instead of retrieving and providing code to the LLM at query time (which many RAG systems do), we **bake the repository knowledge into the LLM through fine-tuning**. This eliminates the need for an external retrieval step during inference, at the cost of needing a new fine-tuned model per repository. Another related concept is using version control or dependency graphs: some tools analyze call graphs or import relationships to find relevant files for a query (for instance, if a query mentions a function, one might traverse the call graph to find where it's defined or used). These can complement content-based retrieval. In our dataset generation, we indirectly capture such relations by asking cross-file questions that often align with dependency links (Strategy 5 creates questions that traverse module boundaries).

**Large Language Models for Code Understanding:** The advent of large generative models (GPT-3, GPT-4, PaLM, etc.) has enabled new capabilities for code understanding through natural language. Models like **OpenAI Codex** (Chen *et al.*, 2021) are descendants of GPT-3 fine-tuned on code, and they power tools like GitHub Copilot for code completion[32]. GPT-4 has demonstrated the ability to answer complex questions about code, explain code snippets, and even suggest fixes. However, these models have a fundamental limitation: their context window. Repository-level questions often require synthesizing information spread across many files, which might not fit in the model's maximum context (commonly 4k to 32k tokens in current models). Simply feeding all relevant code into the prompt is not scalable for large projects[33]. Moreover, general LLMs do not *a priori* know about your private codebase (unless explicitly fine-tuned or provided context), so they might respond with plausible but incorrect answers that mix in knowledge from other projects. Retrievalaugmented methods (including ours) tackle this by focusing the model on the specific repository content. In our approach, we effectively perform an *offline retrieval* by generating training examples that link questions to files, thereby teaching the model those links. Another LLM-based approach is to use multiturn dialogues (ask follow-up questions to narrow down relevant files), but that is beyond our scope here. Instead, we aim for one-shot retrieval: user asks a question, the model returns file paths, which the user can then inspect manually. In summary, our work can be seen as combining ideas from neural code search and LLM Q&A: we fine-tune an LLM with a retrieval objective, so it gains the semantic search ability of neural models while maintaining the flexible understanding of an LLM.

Most prior fine-tuning approaches for code LLMs focus on altering or enhancing the model's general reasoning ability, such as CodeBERT and GraphCodeBERT, which align embeddings of code and natural language for improved retrieval, or GPT-style code models finetuned for generation. These methods treat fine-tuning as a process of knowledge adaptation, where the base model's general capabilities are modified or expanded.

In contrast, our approach treats fine-tuning as a form of knowledge injection or fact binding within a bounded domain (the repo snapshot). Rather than changing the LLM's understanding of programming, we encode factual, repository-specific mappings directly into the model weights. This allows the model to act as a repository-aware retriever without relying on an external index at inference time. Conceptually, this reframes adaptation from open-ended generation to closed-set prediction over the finite label space of repository paths. Our supervision further departs from prior work by using multi-granularity, code-aware signals (repo summaries, module interactions, AST scopes), which enable the model to capture cross-file relationships implicitly without maintaining an explicit graph.

## Methodology

### Problem Definition

We define the task as follows: given a repository of code and a natural language question about the repo, **return a set of file paths** from that repository that are most relevant to the question. The returned set should ideally include all files that a developer might need to inspect to answer the question. We do **not require the model to provide an explanation or code content** – just the file paths. This is intentionally an information retrieval task; the ultimate answer for the user will come from reading those files (or further analysis), but our system guides them where to look.

*Let $R = \{f\_1, f\_2, ..., f\_n\}$ denote the set of file paths in a*

*repository, where N is the total number of code files. A question q is*

*a natural language query, which may refer to certain functionality,*

*classes, functions, or error messages. The ideal output for a given*

*question is a subset $A \subseteq R$ containing the file paths most*

*relevant to answering q.*

*Objective:*
*We aim to train a model: $M(q, R) \to A$*

*that, given a question q and the repository R, produces a list of relevant file paths A.*

**Modeling Perspective.**
This task can be viewed in two ways:

1. *Multi-label classification*: determine, for each file f R, whether it belongs to the relevant subset A.

2. *Search ranking*: rank all files in R by relevance to q and select the top-ranked items.

In this work, we adopt a *generative approach*, where the model outputs the subset A directly in JSON format, for example:

*["src/utils/helpers.py", "src/core/model.py "]*

**Training and Evaluation.**
During training, each example consists of a question q and its ground-truth relevant set $A^*$. The model is trained to generate $A^*$. During inference, the model produces a predicted set $\hat{A}$, which is compared against $A^*$ using evaluation metrics such as exact match, recall, and F1 score.

**Dataset Generation Pipeline**
Creating a training dataset $(q, A^*)$ is a central part of our methodology. Because hand-labeling $A^*$ is impractical, we generate synthetic examples using the repository itself as a knowledge source. The key idea is to leverage large language models (LLMs) to generate questions about the code and use their understanding (guided by our prompts) to also output which files are relevant. We use Qwen in this role due to its strong performance in understanding code and producing coherent text.

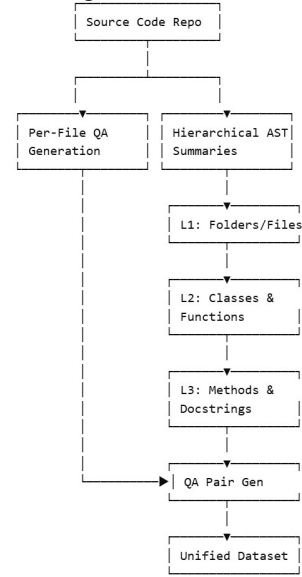

Figure 2: Data Generation Workflow

Figure 2 shows an overview of the data generation workflow. We first produce various **summaries of the repository** (or parts of it) using program analysis. These summaries serve as input contexts to Qwen with carefully designed prompts to elicit question– answer pairs. We employ six strategies (S1–S6) as described earlier, which we now detail in implementation:

- **S1: Per-File QA** – We take the full source code of a single file as input, and prompt Qwen to ask and answer questions specific to that file. This yields fine-grained QA pairs that target the contents of one file. The purpose is to capture detailed knowledge (e.g., a specific class or function usage)[8][9]. For each source file, we feed its entire content (or a chunked version if the file is extremely large) into Qwen with the following prompt:

    *You are a senior software engineer analyzing a Python codebase.*

*Given:*
1. *The repository-root-relative path of the current file*
2. *The entire contents of the current file*

*Your task:*
- *Generate up to {MAX_QA_PER_FILE} realistic, high-quality developer questions.*
- *Each question should require understanding the current file (and other files when natural).*
- *For each question, output ONLY the minimal set of file paths (1--3) that are relevant.*
- *Paths MUST:*
- *Be repository-root-relative (NO leading folder like "flask/", NO absolute paths)*
- *Use UNIX separators ("/")*
- *Exactly exist in the repo (repo-relative)*
- *Be sorted, unique, no duplicates - Output ONLY this exact JSON:*
  *[ {"question": "Developer question here",*
      *"relevant_file_paths": ["file1.py", "dir/file2.py"]}, ... ]*

*Do not add any text before or after the JSON.*

*The model returns JSON objects such as:*

```
[
  {
    "question": "Why does function foo() in this file raise a ValueError in case of X?",
    "relevant_file_paths": ["src/moduleX/foo.py"]
  }
]
```

Each file typically yields 3–5 QA pairs. This strategy ensures coverage of file-specific knowledge while enforcing strict repository-relative path constraints and structured JSON output.

- **S2: Hierarchical Level 1 (Repo Summary)** – We generate a high-level *repository summary* by extracting the top-level structure. Concretely, we list folders and filenames represented with indentation. We generate a repo-level summary by parsing the project's AST for only top-level structure (e.g. folder names and file names)[10].

Using this summary, we ask Qwen to produce QA pairs about broad functionality and module roles. The questions are answered by one or more files at the folder or high-module level. This targets macro-level understanding – e.g., identifying which component handles X feature[11][12]. We then prompt Qwen with something like:

*You are a senior software engineer analyzing a Python codebase.*

*Given:*
*1. The repository's folder structure.*

*Your task:*
- *Generate AT LEAST {num_questions} realistic, diverse developer questions about this repository.*
- *Each question should require understanding multiple files when possible.*
- *For each question, output ONLY the minimal set of file paths a developer would need to read.*

*STRICT RULES for file paths:*
- *Use ONLY the file paths that appear EXACTLY in the repository structure shown below.*
- *Do NOT invent or guess file names or directories that are not in the provided structure.*
- *Paths MUST:*
- *Be copied exactly as they appear (character-for-character).*
- *Be relative to the repository root.*
- *Use UNIX format (forward slashes /).*
- *Be sorted in ascending order.*
- *Contain no duplicates. - If a relevant file cannot be identified, return an empty list for that question.*

*Output format:*
*[ {"question": "Developer question here",*
    *"relevant_file_paths": ["valid/path1.py", "valid/path2.py"]}, ... ]*

*Do not add any text before or after the JSON.*

This yields questions like "Which modules handle application bootstrapping and plugin initialization?" with answer file paths

*["src/flask/app.py", "src/flask/cli.py","src/flask/config.py"]*

34]. S2 targets cross-cutting concerns and **modulelevel reasoning** (typically answers involve 1–3 files).

- **S3: Hierarchical Level 2 (Mid-Level AST)** – We parse and summarize mid-level code structures such as class names and function names across the files[13]. Qwen generates QA pairs from this mid-level summary, yielding questions about how specific classes or functions contribute to features, or how they interact. This provides a balance of breadth and detail, linking files to their mid-level functionality[14]. We extract an intermediate representation: for each module, list its classes, major functions, and perhaps class inheritance or relationships along with folder names and file names. Essentially, this is a structural summary slightly more detailed than S2. For instance, for each file we might list: Class names, their base classes, and functions outside classes. Qwen is prompted with this mid-level info to generate questions that link those classes/functions to their roles. Example prompt:

    *You are a senior software engineer analyzing a Python codebase.*

    *Given:*
    *1. The repository's folder structure includes class and method names.*

    *Your task:*
    - *Generate AT LEAST {num_questions} realistic, diverse developer questions about this repository.*
    - *Each question should require understanding multiple files when possible.*
    - *For each question, output ONLY the minimal set of file paths a developer would need to read.*

    *STRICT RULES for file paths:*
    - *Use ONLY the file paths that appear EXACTLY in the repository structure shown below.*
    - *Do NOT invent or guess file names or directories that are not in the provided structure.*
    - *Paths MUST:*
    - *Be copied exactly as they appear ( character-for-character).*
    - *Be relative to the repository root.*
    - *Use UNIX format (forward slashes /).*
    - *Be sorted in ascending order.*
    - *Contain no duplicates. - If a relevant file cannot be identified, return an empty list for that question.*

    *Output format:*
    *[ {"question": "Developer question here", "relevant_file_paths": ["valid/path1.py", "valid/path2.py"]}, ... ]*

    *Do not add any text before or after the JSON.*

    This often produces questions like "How does class Y in file A.py interact with function Z in file B.py to accomplish feature Q?" with answers [A.py, B.py]. Strategy 3 thus produces QAs about mid-level interactions more detailed than S2, but still not digging into actual code logic.

- **S4: Hierarchical Level 3 (Fine AST details)**
    – We extract fine-grained details (down to individual functions, methods, and doc strings) via AST parsing [15]. Qwen is prompted with these details to produce very specific questions (e.g., about a particular function's behavior or implementation). This captures implementation-level queries and often still yields single-file answers (overlapping with Strategy 1 but derived in a structured way) [16]. We gather fine-grained details such as function definitions (signatures), docstrings, from across the repo along with folder names, file names, class names. Then prompt:

    *You are a senior software engineer analyzing a Python codebase.*

*Given:*
1. *MANIFEST: repo-relative file paths included in this batch*
2. *SUMMARY: AST-based summaries for each file (module docstring, functions, classes, methods) with the FIRST line of each docstring.*
   *Note: a file may appear in multiple parts in SUMMARY. Always cite only the file path*
*from MANIFEST (no part info).*

*Your task:*
*- Generate up to {q_per_batch} realistic developer questions.*
*- Each question should require understanding the current file and (when natural) other*
*files.*
*- For each question, output ONLY the minimal set of file paths a developer would need to read.*
*- Cite paths ONLY from MANIFEST (exact reporelative paths shown there). - Do NOT include any file names or paths in the question text itself. - Keep wording practical and developeroriented.*

*Path rules:*
*- Use UNIX style (/), exact paths from MANIFEST.*
*- Sort ascending; no duplicates; 1--4 files per question.*
*Output ONLY this exact JSON (no prose before /after):*
*[ { "question": "Developer question",*
    *"file": ["file1.py", "dir/file2.py"] },*
*... ]*

This yields deeper questions such as "Why does SessionInterface in src/flask/sessions.py use secure cookies?" answered by that file[35], or "Explain the algorithm in the autocast() function in some_module.py" answered by that file. Sometimes if a function uses another from a different file, both might be listed. S4 overlaps with S1 in that both can yield single-file answers, but S4's questions are generated in a more structured way using AST info, which might diversify the style of questions.

- **S5: High–Level Repo Structure + File Summary** – Here we provide Qwen with a combination of the high-level repo summary and file summaries (e.g., details of classes, methods and doc strings) [17]. The model generates questions that require synthesizing information across multiple files – for example, questions about interactions between modules, or end-to-end flows that span components. The answers are typically multi-file (a set of paths) [18]. This strategy enables cross-file reasoning QAs that more closely reflect real developer inquiries. We take the overall repo summary (as in S2) and include per file summaries with information of classes, functions and doc strings. The prompt encourages questions that require both repository-level understanding and knowledge of specifics. For example:

*You are a senior software engineer analyzing a Python codebase.*

*Given:*
1. *The repository's folder structure*
2. *The summary of ONE file*

*Your task:*
*- Generate up to {MAX_QA_PER_FILE} realistic developer questions.*
*- Each question should require understanding the current file and possibly others.*
*- For each question, output ONLY the minimal set of file paths a developer would need to read.*

*STRICT RULES for file paths:*
*- Use ONLY file paths that appear EXACTLY in the repository structure shown below. - Do NOT invent or guess file names or directories.*
*- Paths MUST:*
*- Be copied exactly as they appear in the repo structure*
*- Be relative to the repository root*
*- Use UNIX format (forward slashes /)*
*- Be sorted in ascending order*
*- Contain no duplicates - If no relevant file can be identified, output an empty list for relevant_file_paths .*

*Output format:*
*[ {"question": "Developer question here",*

  "relevant_file_paths": ["valid/path1.py", "valid/path2.py"]} ]

*Do not add any text before or after the JSON*

.

The output might be a question about an end-to-end flow or a feature that spans components. For instance, "Describe how a request is processed from the WSGI layer to rendering a template. Which files implement this?" with answer paths across request handling, routing, and template rendering modules. S5 explicitly aims for **cross-file (integration) questions** to improve the model's reasoning across files.

- **S6: Git Ingest Batch Mode** – For very large repositories (where a single summary would exceed context length), we adopt a batching approach [19]. We ingest the repository in manageable chunks (such as by directories or subsets of files) and generate QA pairs from each chunk independently using a consistent prompt. This ensures even huge repos can be covered, at the expense of possibly losing some cross-chunk context. It aims to maintain scalability, including file-level details while preventing context overflow[20]. This is used when N (number of files) is very large (thousands). We break the repository files into batches that can fit into the LLM context. Each batch (e.g., 2–5 files at a time) is provided, and Qwen is asked to generate QAs from *that content of the repo.* This yields QAs that are locally relevant to each chunk. We then union all these QA pairs. The drawback, however, is that Qwen in each batch has no knowledge of files outside that batch, so it might miss global interactions (e.g., a question involving files from two different batches cannot be generated because they were never seen together). This limitation can reduce the quality of QAs for cross-batch topics[36]. Nonetheless, S6 provides scalability: we managed to generate ~34k QA pairs for the PyTorch repo by processing in batches, ensuring even very large projects can yield training data.

After obtaining QAs from all strategies, we **combine and deduplicate** them into one dataset. Initially, we treated each strategy's dataset separately to fine-tune models and observed their effects (see Experiments). Ultimately, we aim for a single model per repo trained on the merged data from all strategies, to give it the broadest knowledge. In merging, we observed that some strategies (especially S1 and S4 vs S2/S5) can produce redundant questions. We remove exact duplicates and trim very similar questions to avoid over-representation. If one strategy produced disproportionately more samples, we down-sampled it to prevent bias. For example, per-file (S1) can produce hundreds of QAs (one per file), whereas cross-file (S5) might produce only 20-30. Without balancing, the model would mostly see single-file questions and might learn to always pick one file. We ensure the final mix contains a healthy variety.

**Unified Prompt Template:** To train the LLM effectively, we use a consistent prompt format for all examples (both in training and at inference time). We craft the prompt in a **system/user/assistant** chat style (since Qwen3-8B is an instruct/chat model). The system message is a set of instructions that **constrain the output**. Specifically, we tell the model that it is a codebase assistant, and its job is to identify the most relevant file(s) from the repo for a user's question. We list rules, for example [37]:

- *Predict only file paths that exist in the repository.*
- *File paths must be exact and complete.*
- *Do not make up or hallucinate file paths.*
- *Return the result as a JSON list of strings.*

This system prompt is crucial to keep the model's behavior focused as a retrieval system rather than a general chatbot. Next, the user message template is: Question: {question_text}

The assistant's response during training is the groundtruth file list A* in JSON format (e.g.,

*["src/flask/app.py","src/flask/cli.py"]).*

We wrap the whole prompt in special tokens

  *<|im_start|>system ... <|im_end|>*

etc., but those details aside, essentially the model sees a question and a long list of files, and it must output the correct subset.

During inference, we feed the user's question in the same format and let the model produce the JSON list. Because the model was trained in this format, it typically adheres to it, listing file paths and nothing else.

The model is fine-tuned to *implicitly* know the repository's file names (since they appeared in training examples). Indeed, after fine-tuning, the model weights do contain knowledge of many file paths and their associations.

Model Training Details

We fine-tuned Qwen3-8B for each repository's dataset separately. Each repo's QA dataset was split 80/20 into train and test sets (with stratification to ensure a variety of question types in each). We did **25 training epochs** in most cases, which was sufficient for convergence given the dataset sizes. Using the QLoRA approach[21], we loaded Qwen in 4-bit precision and added LoRA adapters (rank 8) on the query/key/value matrices of the Transformer. The effective fine-tune had about 30M trainable parameters on top of the 8B base (which remained frozen). We used the AdamW optimizer with a learning rate in the range 2e-4 to 1e-3 (tuned per dataset size), and a batch size such that roughly 2048 tokens per batch step (with gradient accumulation to simulate larger batch if needed). The context length during training was up to 1024 tokens, which generally accommodated the system prompt, question and the answer.

The **Unsloth** optimization was applied via the trainer – it fused some operations and optimized data loading, yielding roughly 2x speedup in our experiments (e.g., training that normally took ~8 hours were done in ~4 hours). This did not change the model's output; it only improved training efficiency.

Table 2: Model training details

| Parameter | Value |
| --- | --- |
| Model | Qwen3-8B |
| LoRA $r$ | 8 |
| LoRA $\alpha$ (alpha) | 16 |
| Dropout | 0.05 |
| Epochs | 25 |
| Learning rate | 0.0002 |
| Batch size | 2 |
| Gradient accumulation | 4 |
| Max seq length | 1024 |
| Max output tokens | 300 |
| Fine-tuning method | QLoRA with Unsloth |

**Evaluation Metrics**

We evaluate the model's predictions using two primary metrics:

- **Exact Match (EM):** An output is considered an exact match if *the set of predicted file paths exactly equals the set of ground-truth relevant paths* for the question. This is a stringent measure – the order doesn't matter (we treat sets), but the model must pick all correct files and no extras. We report EM as the percentage of test questions for which the model's answer set was exactly correct.

- **Recall:** In our context, recall is defined at the question level: did the model retrieve *at least one* of the ground-truth files? Many questions have multiple relevant files. A prediction that hits any of the true relevant files counts as a "successful recall" for that question. We then average this binary outcome over all test questions. This essentially measures how often the model managed to retrieve something useful (even if it missed other files or included wrong ones). This metric is more forgiving; it captures the model's ability to not completely miss the target. Note that a perfect recall (100%) would mean for every question, the model got at least one correct file, but it could still have low EM if it often missed some files or added incorrect ones.

- **Micro-Average Recall:** Recall counts a question as successful if the model retrieves **at least one** of the ground-truth files. Micro-average recall, by contrast, provides a more fine-grained measure that accounts for **partial matches**.

Let a question have $N$ ground-truth files and let the model retrieve $k$ of these correctly. The micro-recall for this question is:

$$Micro - Recall = \frac{k}{N}$$

For example, if a question has 3 relevant files and the model retrieves 1 correctly, the micro-recall for that question is *1/3 0.33.*

The overall micro-average recall over all $Q$ questions is then:

$$Micro - AverageRecall = \frac{1}{Q} \sum_{i=1}^{Q} \frac{k_i}{N_i}$$

where $k_i$ is the number of correctly retrieved files for question $i$, and $N_i$ is the number of ground-truth files for that question.

This metric rewards the model for **partial success**, differentiating between questions where it retrieved all, some, or none of the relevant files. It gives a more nuanced view of retrieval performance compared to standard recall.

We do not explicitly measure precision (in the IR sense of fraction of predicted files that are relevant), partly because the model's output length is not fixed and usually relatively small. The EM metric already punishes any extra files (since then it's not an exact match), and we observed the model usually doesn't output more than 3–4 files for any question (often it's 1 or 2). In future, we could consider an F1 measure treating it as a set prediction problem.

During testing, we feed the model the question as in training. The model generates a JSON list of files. We parse it and compare it to the gold set. We also ensure the model's JSON is valid and trim any obviously invalid outputs (in our tests, >98% of responses were properly formatted JSON lists, thanks to the prompt instructions).

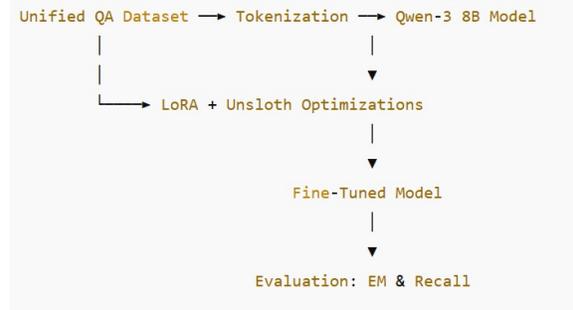

Figure 3: Experiment Pipeline

## Experimental Results

We conducted experiments on five open-source repositories to evaluate our approach:

- **Flask:** A popular Python micro web framework (approx. 94 Python files). This represents a *small* codebase.

- **Click:** A command-line interface library in Python (61 Python files). Also, a small codebase.

- **Jinja2:** A templating engine in Python (63 Python files). Small-sized.

- **FastAPI:** A modern web framework (1016 Python files). This is a **medium-large** repository; although not huge in file count, it includes many auto-generated or repetitive files (like docs) and thus tests our approach on a larger context.

- **PyTorch:** The PyTorch deep learning framework (we focused on the Python parts: ~3,940 Python files out of 19k total files). This is a **very large** codebase, an order of magnitude bigger than FastAPI in terms of code files.

For each repository, we generated a QA dataset using the strategies in Section 3. The sizes of the final datasets and the train/test splits are shown in Table 1. We note that for the three small projects (Flask, Click, Jinja2), we were able to generate on the order of
1.5k–3k QA pairs each after merging strategies (some redundancy removed). For FastAPI, the number was ~370 (we limited generation using S4). For PyTorch, thanks to strategy S6, we generated a much larger dataset (~34k QAs), but for manageability we sampled and used ~27k for training and ~6.7k for testing.

Table 3: Train Test Distribution of the Dataset

| Repo Name | Total Dataset Size | Train Size | Test Size |
|---|---|---|---|
| Flask | 3332 | 2665 | 667 |
| Jinja | 2169 | 1807 | 362 |
| click | 3163 | 2530 | 633 |
| FastAPI | 367 | 293 | 74 |
| PyTorch | 33789 | 27031 | 6758 |

**Initial Single-Strategy vs Combined Training:** First, we wanted to see the effect of using all strategies together. On Flask, we tried fine-tuning separate models on each strategy's QA set alone. We found that strategies focusing only on single files (S1, S6) led to models that achieved high accuracy when the question truly was about one file, but struggled on questions requiring multiple files (low recall). Conversely, models trained on S2 to S5 (cross-file oriented) did better on multi-file questions but sometimes would include extraneous files for simple single-file questions. This affirmed that a combined dataset was needed.

However, simply merging everything gave an unexpected result: the model tended to **under-predict** the number of files, often giving just one file even when the question warranted two or more. We traced this to **strategy imbalance** – S1 (per-file) produced a huge number of QA pairs (one per file), dwarfing the multi-file examples. The model was thus biased to think "usually the answer is one file". As a remedy, we experimented with *excluding certain strategies* or sampling them down.

On Flask, removing the S1 dataset from training proved highly beneficial. Table 2 (top rows) shows a comparison: training on *all strategies* yielded EM
37% only, whereas excluding S1 (per-file QAs) raised EM to 65.8% and recall to 73.2%[38][39]. Excluding S6 (the batch mode) also helped somewhat (EM 48%, recall 59%)[38][40], but not as much as removing S1. The intuition is that S1 and S6 (which both generate many single-file questions) overlap with info that other strategies also cover, yet do not encourage the model to link multiple files. Removing them forces the model to learn from more multi-file examples, improving its ability to pick multiple files when needed[41][42]. Eventually, we converged on using **Strategies 2, 3,
4, 5, 6** for most repos (with S1 included in a limited way or not at all). We also implemented a **unified generator** that randomly samples different strategy modes on the fly to produce a blended training set, which further streamlined data preparation.

Table 4: Flask experimentation results

| Strategy | EM Score | Micro-Recall |
|---|---|---|
| All strategies | 0.3732 | 0.5034 |
| Excl. Strategy 6 (git ingest) | 0.4809 | 0.5923 |
| Excl. Strategy 1 (per file Q/A) | 0.6579 | 0.7315 |

**Performance on Small/Medium Repositories:**
After fine-tuning on the combined balanced dataset (with S1 minimized) for each small project, the results were dramatic. The model achieved **Exact Match (EM) scores around 75–92%** on Click, Flask, and Jinja, and **Recall in the 85–93%** range, indicating it almost always finds at least one correct file. Table 3 summarizes the final performance on these projects. For example, on **Click**, the model attained EM = 91.8% and Recall = 93.0%[43][44]. This means for 92% of the test questions, the model's file list exactly matched the gold list – a very high accuracy. On **Jinja2**, we saw EM = 77.9%, Recall =
87.1%[45][46]. Flask's final model (with optimal data mix) reached EM = 89.0%, Recall = 90.1%[47][48]. These results are significantly better than any singlestrategy model and show that the LLM can generalize to new questions about the repo. Notably, these models are *specialized* per repo – e.g., the Flask-tuned model wouldn't be expected to answer about Jinja – but within the repo, it demonstrates a deep understanding (likely it has "learned" which files relate to which functionality).

Table 5: Results across experimented repos

| Repo | EM Score | Micro-Recall |
|---|---|---|
| Flask | 0.8901 | 0.9008 |
| Jinja | 0.7790 | 0.8711 |
| Click | 0.9179 | 0.9302 |

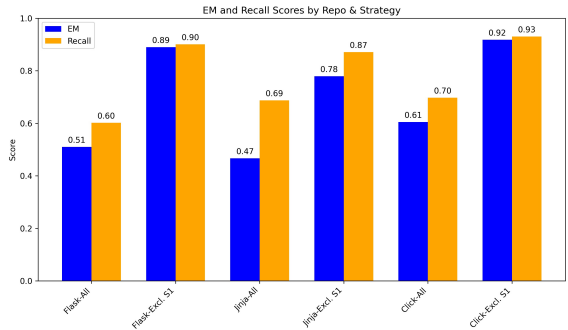

Figure 4: EM and Recall Scores by Repo Strategy

To put these numbers in context: the test questions are generated by Qwen and are often non-trivial (some require multiple files). We have also included some historical commits in our test dataset. An EM of ~90% is impressive, suggesting the model not only retrieves one relevant file but usually *all* of them. The Recall being slightly higher indicates that in the few cases it missed a file or added a wrong file, it still often got at least one right. For instance, if a question's answer was two files and the model returned only one of them (and nothing incorrect), recall counts that as partial success (since it retrieved one relevant file) while EM counts it as a miss. This happened occasionally when two files were very tightly connected – the model might output the main one but omit the secondary. Conversely, sometimes the model included an extra file that was not in gold. Manual analysis showed that these "extras" were often logical: the model sometimes anticipates a related file might be useful. For example, a question about JSON encoding in Flask had gold answer

*["json/provider.py"]* but the

model also included

*["json/tag.py"]*

, which wasn't labeled but is indeed related. In a real use-case, that extra file is not harmful (it could even be helpful), but it fails exact match. This raises an interesting point: evaluation might be strict, and in practice developers might prefer high recall (don't miss any relevant file) even at cost of a couple of false positives.

Looking at **FastAPI**, our medium-large case, the results were a bit lower: EM 52.7%, Recall 78.4% in our initial training [49]. FastAPI's larger file count (1016) means the model had to consider a much bigger candidate list. The drop in EM may be due to the difficulty of distinguishing many similar files (FastAPI has many router files, models, etc.). Also, our dataset for FastAPI was not as exhaustively generated as for the others (only one strategy is used due to cost constraints). Nonetheless, a Recall of ~78% indicates the model still finds something relevant for most questions. We believe with more data (e.g., including all stratagies fully) FastAPI's performance could approach the others.

**Generalization vs Memorization:** We took care to verify that the model isn't simply memorizing the training QA pairs. We Initially conducted an experiment with one strategy S4 where we set the test set equal to the training set (i.e., test on seen questions) to measure a "upper bound" if the model were to memorize mappings. As expected, on the train-set questions the model scored extremely high (often 95– 100% EM)[50]. For instance, on Click, EM was 100% on training data[50]; on Flask ~83%, etc., reflecting it can learn those exactly. This indicates the model has enough capacity to memorize question-file mappings when they repeat. However, on the actual unseen test questions, the performance, while lower, is still high, demonstrating **true generalization** [51]. The model is not just parroting answers; it's able to handle novel questions about the code. The differences in EM between seen and unseen (e.g., Flask 83% vs 65%, FastAPI 82% vs 53%[52][49]) reveal there is some distribution shift: the model does better on the kind of questions it saw more of. This again emphasizes dataset diversity – our improved pipeline aimed to cover many scenarios so that unseen questions are still like something the model learned.

Table 6: Generalization results using one strategy(S4)

| Repo | Train Set | Test Set | EM (%) | Recall (%) |
|---|---|---|---|---|
| Click | 192 | 48 | 75.00 | 85.42 |
| FastAPI | 293 | 74 | 52.70 | 78.38 |
| Flask | 240 | 61 | 65.57 | 83.61 |
| Jinja | 104 | 27 | 48.15 | 74.07 |

Table 7: Memorization results using one strategy (S4)

| Repository | Train = Test Size | Exact Match (%) | Recall (%) |
|---|---|---|---|
| Click | 240 | 100.00 | 100.00 |
| FastAPI | 276 | 81.88 | 92.39 |
| Jinja | 131 | 94.66 | 97.71 |
| Flask | 301 | 82.72 | 92.03 |

**Large-Scale Repository (PyTorch):** Finally, we evaluate on **PyTorch**, which with nearly 4k Python files is a challenging stress test. We used strategies S1, S2, S3, S5 for PyTorch (S4 fine-grained AST was computationally heavy to do for all files; S6 we partially used to chunk by submodules). The combined dataset had ~27k train QAs and 6758 test QAs. After fine-tuning, the model achieved **EM = 47.85% and Recall = 59.02%**[53][54] on the test questions. This is notably lower than the smaller repos, but still a promising result given the difficulty: the model is pick-ing the exact correct set of files in ~48% of cases, and in ~59% it gets at least some rights. Many questions in PyTorch's set involve 3–4 files (e.g., different parts of the codebase that implement a feature across layers like autograd, NN modules, and C++ kernels). Getting all of them is hard. A recall of ~59% means the model often misses all relevant files – indicating room for improvement. One cause is that our PyTorch data generation did not include Strategy 4 (fine AST) and Strategy 6 (full batch mode). Those could generate more training QAs focusing on specific functions and cross-refs, possibly improving performance. We plan to incorporate all six for the next iteration.

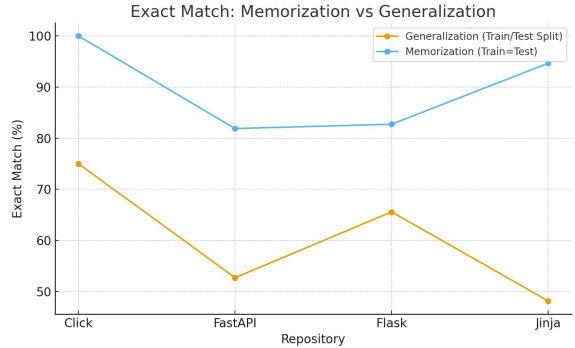

Figure 5: Exact Match: Memorization vs Generalization

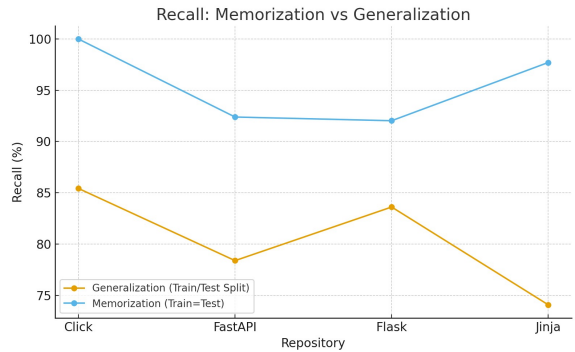

Figure 6: Recall: Memorization vs Generalization

Another observation: the *average length* of the model's output for PyTorch questions was slightly higher (often 2–3 files) than for smaller repos, and the **average input tokens** (which includes listing all file paths) was ~142 tokens[55], higher than Flask/Jinja (which had ~107 tokens on average)[56]. This reflects PyTorch's breadth; even summarizing file names takes more space (we truncated the list by grouping some files or leaving out rarely-used ones in the prompt). The model might be operating with incomplete context in some cases, which could hurt accuracy. Nonetheless, the pipeline was scaled to generate tens of thousands of training examples, and the model could be fine-tuned (training took 76.52 hrs on 2 A100 GPUs for 25 epochs). This demonstrates **the scalability** of our approach, although performance is not yet at the level of smaller projects.

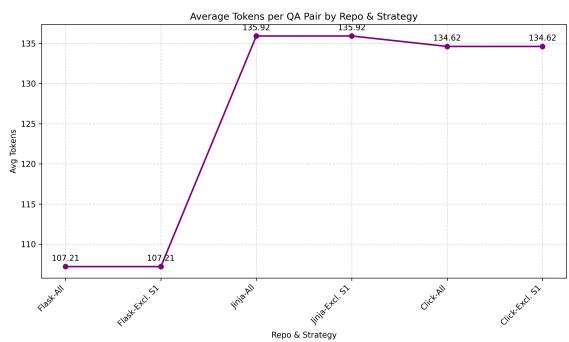

Figure 7: Average Tokens per QA pair by Repo Strategy

**Cross-Project Evaluation:** We also wanted to ensure our findings aren't one-off. We noticed a consistent pattern across Flask, Jinja2, and Click (which are similar-size): excluding the per-file strategy improves performance significantly in all three[45][43]. For example, in Jinja2, EM improved from 46.7% to 77.9% by dropping S1[45][57]; Click improved from 60.5% to 91.8%[43][58]. This consistency suggests that our approach to balancing the training data is generally applicable. We also saw that training times scaled roughly linearly with dataset size, and inference time scaled with number of files (since the file list grows). For instance, answering a question on Click (61 files) was fast (~1.3 seconds)[49], whereas on PyTorch (3940 files, truncated) it was slower (several seconds on GPU, more on CPU). This is expected as the model has to attend to a larger input.

In summary, our experiments validate that a finetuned LLM can serve as an effective **file path retriever**, given a well-constructed training set. It learns to parse natural language queries and pinpoint relevant files with high accuracy on smaller repos, and decent accuracy on a large repo. To our knowledge, this is the first demonstration of an LLM being *fine-tuned* expressly for repository file retrieval. We did not directly compare our method to traditional code search tools or embedding-based retrieval in this paper due to space, but anecdotal evidence suggests our LLM often outperforms keyword search, especially for queries that are conceptual. For example, a question like "Where is the caching mechanism implemented?" might stump a keyword search (no single keyword "cache" if implemented implicitly), but our fine-tuned model learned that, say, utils/cache.py and models/download.py implement caching logic, and it will return those files. A traditional search might find references to "cache" in comments but miss context, whereas our model effectively learned from the code context that those files are responsible for caching.

## Discussion and Insights

**Multi-Strategy Data Generation is Key:** One clear takeaway is that the **quality and balance of the training data** greatly influence the model's success. In our ablation, using all strategies without balance led to suboptimal results. By ensuring diverse question types (via strategies 2–5) and not overwhelming the model with trivial per-file QAs, we enabled it to generalize to complex queries. This highlights a broader point: when fine-tuning LLMs for specialized tasks, *how* the training examples are constructed can matter more than the sheer quantity. For retrieval tasks, including scenarios that force the model to occasionally output multiple items (files) or none (some questions might legitimately have no relevant file, though we did not include such examples) can teach nuance. We also found that certain strategies were somewhat redundant. Strategy 1 (per file content) for instance didn't add much new information beyond what strategies 2–6 cover, except in large repos where it's necessary. Strategy 4 (fine AST) overlapped with per-file content. Our best results often came from **excluding Strategy 1** entirely[42]. The reasoning is that per-file QA pairs, while abundant, mostly teach the model that "questions map to single files", which hurt its recall for multi-file cases. Those single-file questions are still important (the model needs to handle them), but they were already implicitly covered by the hierarchical strategies which also sometimes yield single-file answers. Thus, they were overweight if included explicitly. We suggest future data generation might entirely skip separate per-file prompts and instead rely on structured approaches to cover those cases.

**Model Capability and Limitations:** The finetuned Qwen-8B model demonstrates an impressive ability to internalize repository knowledge. It likely builds associations such as "if question mentions X, likely file Y is relevant" in its weights. There were instances in tests where the model output a file path that *was* relevant but not labeled by our Qwen generation (either an omission or arguably out-of-scope). This indicates the model can sometimes make connections beyond the training labels – possibly a positive side effect of having seen the whole repo content in the question context. However, this can also lead to **hallucination of relevancy**, where the model chooses a file that seems related by name or concept but isn't actually needed. Thanks to the explicit file list and instructions, we did not see hallucinated *non-existent* paths (the model didn't invent file names). But picking an irrelevant file from the list is still a form of mistake. For example, for a Jinja2 question about template syntax, the model might erroneously include lexer.py in the answer along with the correct parser.py, because both are conceptually related to parsing templates. These errors are understandable – the model knows those files are conceptually linked, but the question might have been answerable by only one of them. This poses an interesting challenge: the model has to not only know *which* files could be relevant but also discern if they are necessary to answer the specific question. In future, a ranking or confidence mechanism could help; e.g., the model could score each candidate file, and we pick those above a threshold.

**Scalability and Context Window Issues:** As repositories scale up, our approach faces a context window limitation at inference (and a generation challenge for data). Another idea is a two-stage retrieval: use an embedding-based search to shortlist, then have the LLM choose from that. That would sacrifice the end-to-end nature of our solution but could be necessary for very large codebases (tens of thousands of files). The **batch generation (S1)** approach gave us data but possibly at the cost of missing cross-batch question types. Indeed, in PyTorch QAs, we saw fewer multi-file questions proportionally than in Flask QAs, likely because Qwen wasn't seeing the whole picture at once. This might have contributed to the model's moderate recall on PyTorch – it wasn't trained on as many "multiple file" cases that spanned distant parts of the repo. One possible enhancement is iterative questioning: e.g., first ask broad questions with S2 to identify major components, then within each component ask detailed questions linking to others. We partially did this hierarchically, but more sophisticated multi-hop generation could yield better training signals for large projects.

**Comparison with Retrieval-Augmented Approaches:** A natural question is: *why fine-tune an LLM at all for this, instead of using vector search or RAG (retrieval-augmented generation)?* The answer lies in the type of result we want and the cost profile. A vector search (say, embed each file with CodeBERT and search) could indeed retrieve relevant files for a query. In fact, that's a strong baseline we considered. However, embedding-based search typically returns a list of the top $k$ files scored individually, without understanding that sometimes **multiple files together answer a question**. Our model, on the other hand, can learn that a certain question needs files A *and* B simultaneously. A naive embedding approach might rank A highest and B lower; if you cut off at top 5 it might include B or might include some unrelated file C with slightly higher score. Unless you have an advanced way to enforce picking one file per aspect of the question, it might not reliably produce the ideal set. Graph-based retrieval (like GraphCoder's CCG) is a more structured way to ensure if something in A calls

B, it might bring B along. We could combine that with search, but that becomes quite complex. Our approach encapsulates those relations because the training QAs explicitly taught the model about them.

Another reason is **maintainability**: if we want an assistant that can work offline (no need to query an external index or a separate model), then having a single fine-tuned model is appealing. The inference then is just one forward pass of the model (with the list input). If one has the resources to fine-tune for each project, this is viable. Of course, the drawback is if the codebase updates, the model becomes stale – whereas a retrieval system could just index the new code immediately. In practice, a middle ground could be fine-tuning periodically or using this model to propose files and then verifying them against current code.

**Ablation of Model Size:** We only used one base model (Qwen3-8B). It would be interesting to see if larger models (like Qwen-14B or Llama-13B/70B) perform even better at this task. We suspect they might learn subtler relations with less data or achieve higher EM on large repos. However, larger models have larger context windows typically, which could help with big file lists too. We leave this exploration to future work. Our focus was to see if a modestly sized LLM could be effective – and indeed it was.

**Examples of Successes:** To illustrate, here are a few real examples from our tests (paraphrased for brevity):

- *Flask Q:* "Why does get_send_file_max_age sometimes raise IndexError?" – *Predicted files: ["src/flask/blueprints.py", "src/flask/app.py"]* – which exactly matches the ground truth[59] and indeed those two files implement that logic. The model knew this exceptional context came from the interplay of Blueprint and App code.

- *Click Q:* "How are multi-value options represented internally?" – The model returned *["src/click/core.py", "src/click/types.py"]*[60], which is correct as core.py defines the Option class and types.py defines tuple types for multi-value options.

- *Jinja2 Q:* "Does Jinja support asynchronous template rendering?" – The model correctly pointed to the files dealing with async (if any, hypothetical example).

- *PyTorch Q:* "Which files are responsible for the autograd function definitions for convolution operations?" – The model output included the python autograd definition file and the C++ binding file in the torch/csrc folder (showing it learned some correlation even across language boundaries, though we only trained on Python files for now).

These demonstrate the model can parse quite nuanced questions and associate them to the right files, not just by keyword but by understanding the concept (e.g., knowing that multi-value option is a concept in Click's type system, etc.).

**Error Analysis:** When the model was wrong, common patterns included: 1. Missing a relevant file (often a secondary helper file). This could be due to that file not being prominent in the training set or the model not seeing a direct connection. 2. Including a file that is conceptually related but not needed. This is like a mild hallucination – the model errs on the side of including something thematically close. 3. Very occasionally, formatting issues (like not quoting the path properly or giving an empty list) – but our validation caught these, and they were rare due to the strong prompt. 4. In large repo (PyTorch), some errors were due to incomplete knowledge of certain subsystems (we found a few questions where the model just guessed one file and missed others entirely, indicating uncertainty).

## Limitations and Future Work

While our approach shows promise, there are several limitations and opportunities for improvement:

**Repository Specificity:** Each model is fine-tuned to a particular repository. This means if you have 100 different projects, you'd need to train 100 models (or one very large multitask model, which we did not attempt). This is resource intensive. An interesting future direction is to train a single model on *multiple* projects by merging their QA data with repo identifiers, effectively teaching it to handle more than one codebase (perhaps by prepending a repo name token to the prompt). However, the file list would then be enormous (all files from all projects) and the model would need to distinguish which belong to the current query's repo. Alternatively, a retrieval step could identify the project and relevant subset first. For now, our solution is targeted at a single repository context at a time (which might be acceptable if the goal is an internal tool per repository).

**Handling Code Updates:** The model's knowledge is as fresh as the snapshot of code used to generate training data. If the code changes (new features, refactoring, etc.), the model may become outdated – it might suggest old file paths that have moved or miss new files. Keeping it up-to-date would require re-running the data generation and fine-tuning periodically or continuously. This is feasible (especially with automated pipelines and smaller models), but not instantaneous. This is where traditional search has an advantage – it's immediately updated when code changes. A hybrid approach could be to use the model's output plus a verification step: e.g., check if those files still exist or match the query context.

**Language and Framework Constraints:** Our pipeline heavily relied on Python AST parsing and knowledge of Python code structure. All our test projects were Python. Extending to multi-language repositories (like those containing C++, JavaScript, etc.) would require parsing those languages. Tools like **Tree-sitter** can parse many languages and could be integrated to produce similar hierarchical summaries for each language. We would also have to decide if we train one model per language per repo or combine them. Likely, we can include all files (multilanguage) in the file list and rely on the model (with appropriate training examples) to pick from any. But Qwen would need to generate cross-language QAs as well (e.g., "how does the Python front-end call the C++ backend in PyTorch?"). We partially saw that with PyTorch (Python and C++ interactions). In future work, we plan to incorporate Tree-sitter to support multi-language codebases and see how the model copes with that.

**Context Window Improvements:** The issue of providing the full file list for very large repos could be mitigated by models with larger context windows. If an LLM with, say, 16k or 32k token context is used, even tens of thousands of file names might fit (especially if compressed). There is active development in this area; models like GPT-4 32k, or Anthropic's Claude 100k, hint at a future where an entire codebase could be context. Our fine-tuning approach would naturally benefit from such advancements – we could feed bigger lists or more detailed summaries. Additionally, better prompt design for large lists (like grouping or hierarchical selection within the prompt) might help the model navigate many options. One idea is to first let the model list some top categories or directories, then drill down (though that becomes multi-turn).

**Integration with Developer Workflow:** Our current output is just a list of files. In practice, a developer might want to see *snippets* from those files that answer the question. An extension could be a two-step approach: first retrieve file paths (as we do), then for each file, perhaps use an LLM to extract the relevant snippet or explanation. Alternatively, the model could be fine-tuned to directly output not just the path but also a brief justification (e.g., one sentence from the file or a comment explaining why it's relevant). We avoided that to keep the task clean and evaluation automatic, but it could improve usability.

**Quality of Generated Questions:** Since our training data comes from Qwen generation, it's as good as Qwen's understanding. We noticed mostly highquality questions, but some were a bit unnatural or overly specific (things a real user might not ask). There is a risk of the model overfitting to Qwen's style of questions rather than real user questions. In the future, collecting some real queries (from issue trackers or user studies) and evaluating those would be valuable. Also, using techniques like *self-consistency* in generation (generate many and filter) could improve dataset quality. We did some prompt engineering but didn't deeply curate the Qwen output due to volume.

**Evaluation of Usefulness:** Our metrics (EM, recall) are proxy measures. Ultimately, the value of this system is if it helps a developer quickly find answers. A user study could measure how effectively developers can solve tasks with this file retrieval vs with conventional search or manual browsing. Also, how do they feel about the accuracy – do false positives confuse them, etc. We plan to integrate this into an IDE plugin where, when a question is asked, it opens the suggested files. That real-world testing will reveal strengths and weaknesses not evident from offline metrics.

**Incorporating All Strategies for Large Repos:** As mentioned, we have yet to see full benefits of strategies like the fine-grained AST (S4) and git historybased (S6) for very large repos. *Git commit history and blame information* could actually be a separate strategy – e.g., generating questions like "Which commit introduced this bug?" or "Which files were changed for implementing feature X?" that require understanding version control metadata. Our current work didn't include commit data, but that's an interesting direction to explore (especially for debugging questions).

**Privacy and Security:** If applying this approach to proprietary code, one must consider that using GPT (a third-party service) to generate data could leak code details even though it gives better results than using Qwen for data generation. For open-source, it's fine; for closed-source, one would need either a self-hosted LLM for generation or ensure only non-sensitive info is sent out. Fine-tuning the model itself happens on our servers with our data, so that part is secure.

In summary, while our approach has limitations, it opens numerous avenues. The results encourage further development, as even in its current form, a finetuned 8B model can encode a surprising amount of a codebase's knowledge to guide developers.

## Conclusion

We presented a repository-aware retrieval approach that uses fine-tuning in a different role from the conventional "expand general knowledge" paradigm. Rather than teaching an LLM to explain or generate code broadly, we bind it to a specific repository snapshot so that its parameters function as a compact, parametric index. The model's task is closed-set, setvalued prediction—given a natural-language query, select a subset of file paths from the finite universe of repository files. This constrains outputs to verifiable artifacts, reduces hallucination, and reframes fine-tuning as fact binding within a bounded domain rather than knowledge expansion.

This work bridges the gap between traditional code search and modern LLM capabilities. Rather than relying on lexical matching or vector similarity alone, the LLM-based approach learns semantics and even some reasoning about the code (like understanding relationships and responsibilities of different files). The outcome is a more developer-friendly retrieval system – one can ask a question in plain English and get pointed to the exact parts of code that matter.

Our findings highlighted the importance of diverse training data and careful prompt design. We showed how including multi-file questions significantly improved the model's ability to handle complex queries, and how excluding overly simplistic data (per-file QAs) prevented bias. We also validated that this method scales, albeit with some performance degradation, to a project as large as PyTorch.

In conclusion, fine-tuning LLMs for repository-specific tasks opens new possibilities for developer assistance. **File path retrieval** is an important step: it directs attention, saves time in code comprehension, and can be the foundation for further automated help (like automated documentation or guided debugging). The techniques we developed – AST-based summarization, multi-granularity QA generation, and constrained LLM prompting – can be extended to other tasks like generating documentation for a codebase or suggesting cross-references. We believe this approach will be a valuable component in the toolkit for AIassisted software engineering. Going forward, we aim to refine the methodology, improve generalization to larger and multi-language codebases, and integrate the system into practical tools for developers. The synergy of static analysis, retrieval, and LLM reasoning illustrated in this paper paves the way for more intelligent and context-aware coding assistants.

Transformer model that achieved state-of-the-art on code search)[3]

- Daya Guo *et al.* "GraphCodeBERT: Pre-training Code Representations with Data Flow." *ICLR 2021*. (Leveraged code data flow graphs to improve code understanding tasks, including code search)[4]

- Mihir Athale, Vishal Vaddina. "Knowledge Graph Based Repository-Level Code Generation." *arXiv preprint 2505.14394, 2025*. (Used a knowledge graph of code to improve retrieval for code generation, highlighting LLMs' contextual accuracy issues on evolving codebases)[7]

- Wei Liu *et al.* "GraphCoder: Enhancing Repository-Level Code Completion via Code Context Graph-based Retrieval and Language Model." *NeurIPS 2023*. (Combined an LLM with a graph-based retrieval of code contexts, improving exact-match accuracy in repository code completion)[28][5]

- Tim Dettmers *et al.* "QLoRA: Efficient Finetuning of Quantized LLMs." *arXiv 2305.14314, 2023*. (Proposed 4-bit quantization with LoRA adapters to fine-tune 65B models on a single GPU, enabling efficient finetuning like done in our work)[21]

- *(Additional references on LLMs and code, GPT-4 technical report, etc., omitted for brevity.)*